\documentclass[a4paper,fleqn,usenatbib]{mnras}

\usepackage{newtxtext,newtxmath}

\usepackage[english]{babel}
\usepackage[dvipsnames]{xcolor}

\usepackage[T1]{fontenc}
\usepackage{ae,aecompl}

\usepackage{graphicx}	
\usepackage{amsmath}	

\title[Electric field anisotropy in solar wind]{Scale dependent anisotropy of electric field fluctuations in solar wind turbulence}

\author[Deepali, Supratik Banerjee]{
Deepali,$^{1, 2}$\thanks{E-mail: ddeepali@hs.uni-hamburg.de}
Supratik Banerjee$^{2}$\thanks{E-mail: sbanerjee@iitk.ac.in}\\
$^{1}$Hamburg Observatory, University of Hamburg, Gojenbergsweg 112, 21029, Hamburg, Germany\\
$^{2}$Department of Physics, Indian Institute of Technology Kanpur, 208016, Uttar Pradesh, India\\
}

\date{Accepted XXX. Received YYY; in original form ZZZ}

\pubyear{2020}

\begin{document}
\label{firstpage}
\pagerange{\pageref{firstpage}--\pageref{lastpage}}
\maketitle

\begin{abstract}
We study the variation of average powers and spectral indices of electric field fluctuations with respect to the angle between average flow direction and the mean magnetic field in solar wind turbulence. Cluster spacecraft data from the years 2002 and 2007 are used for the present analysis. We perform a scale dependent study with respect to the local mean magnetic field using wavelet analysis technique. Prominent anisotropies are found for both the spectral index and power levels of the electric power spectra. Similar to the magnetic field fluctuations, the parallel (or antiparallel) electric fluctuation spectrum is found to be steeper than the perpendicular spectrum. However the parallel (or antiparallel) electric power is found to be greater than the perpendicular one. Below 0.1 Hz, the slope of the parallel electric power spectra deviates substantially from that of the total magnetic power spectra, supporting the existence of Alfv\'enic turbulence.  
\end{abstract}

\begin{keywords}
(Sun:) solar wind -- (magnetohydrodynamics) MHD -- turbulence -- methods: data analysis
\end{keywords}



\section{Introduction}
The solar wind provides a natural laboratory to study space plasma turbulence which is believed to be responsible for its acceleration and heating. Despite being weakly collisional in nature, turbulence in the solar wind plasma can be studied in the framework of magnetohydrodynamics (MHD) for the length scales superior to the ion gyroscale and ion inertial scale ($\sim 100$ km). Similar to neutral fluids, homogeneous MHD turbulence also entails an inertial range energy cascade with a scale invariant flux rate $\varepsilon$. This holds good only for relatively short intervals of solar wind where the assumption of homogeneity can be justified. From spacecraft, the turbulent fluctuations of the plasma fluid and the electromagnetic field are directly obtained in the form of a time series. Since both the spacecraft speed ($\sim$ 1 km/s) and the Alfv\'en speed ($\sim$ 50 km/s) of the solar wind at 1 AU are much lower than the solar wind speed ($v \sim$ 600 km/s), one can use Taylor's hypothesis to map length scales ($\ell$) into corresponding time scales ($\tau$) as $ \ell = v \tau$ or different frequencies ($f$) to equivalent wave numbers ($k$) as $ \omega = v k$ \citep{Taylor, banerjeethese}. The power spectra for the magnetic field, velocity and density fluctuations in the solar wind have been extensively studied for many decades. In the MHD frequency range of $10^{-4}-10^{-1} Hz$, where both the magnetic energy and density power spectrum are found to scale as $f^{-5/3}$ (or $\sim k^{-5/3}$), velocity power spectrum is found to follow a shallower $f^{-3/2}$ spectrum \citep{Podesta2006}. Turbulence power spectrum for higher frequencies are also investigated systematically \citep{Alexandrova2009, Sahraoui2009}. It is only recently that the first measured power spectrum of electric fluctuations has been reported \citep{Bale}. The spectrum is shown to have a power law index of $-5/3$ and follows the magnetic field fluctuation spectrum until around 0.45 Hz (breakpoint) at the spacecraft measured frequencies. Beyond this breakpoint, the magnetic spectrum becomes steeper whereas the electric spectra becomes enhanced.

All these power spectra have been obtained under the assumption of isotropy. In MHD, unlike velocity, Galilean transformation cannot eliminate the effect of mean magnetic field ($B_0$) which leads to unequal deformation of the Alfv\'enic wave packets along and perpendicular to $B_0$. Due to a non-negligible $B_0$, anisotropy in solar wind turbulence (SWT) is normally expected in the turbulence power level as well as in the spectral indices. However, in studies conducted using a global mean magnetic field, only power anisotropy and no spectral index anisotropy were reported \citep{Tessein, Sari, Chen2011}. This puzzle can be addressed using the results of several numerical studies of MHD turbulence showing that the deformation of wave packets of a specific length scale will be regulated by the mean magnetic field of comparable length scale rather than that of the largest scale \citep{Cho2000, Milano2001}. For solar wind, the minimum variance direction of the inertial range fluctuations is found to closely follow the scale dependent local mean field (LMF) \citep{Horbury95}. It is therefore reasonable to investigate the power and spectral index anisotropies of inertial range solar wind turbulence with respect to the scale dependent mean magnetic field rather than the global mean field. For single spacecraft data (in the form of a one dimensional time series), an estimate of the LMF and the analysis of anisotropy with respect to the LMF are performed using the method of wavelets. Unlike the previous studies with global mean field, here, prominent power and spectral index anisotropies of magnetic power spectra are reported both for polar and ecliptic solar winds \citep{Horbury, Podesta2009}. Within the frequency range $10^{-2}-10^{-1}$ Hz, the parallel power is found to be at least 5 times less than the perpendicular power whereas the spectral index varies from around -2 to -5/3 as the angle between the LMF and the flow direction changes from $0^\circ$ to $90^\circ$, being consistent with the critically balanced MHD turbulence theory by \citet{Goldreich}. Clear signatures of anisotropy for (i) magnetic power spectra at MHD and sub-ion scales and (ii) inertial range velocity spectra are also found for near ecliptic solar wind using both single and multi spacecraft data \citep{Chen2010, Chen2011, Wicks11}. Recently, it is realized that the scale dependent spectral anisotropy may indeed emerge from the mixing of the scaling property of the LMF and that of the magnetic field fluctuations \citep{Oughton}. This effectively makes the magnetic power spectra to be derived from moments of order higher than 2 thereby leading to a sensitivity towards the phase randomization. The signature of critical balance cannot therefore be ascertained just by the spectral anisotropy with respect to an LMF. In fact using Hilbert spectral analysis of solar wind data, both the parallel and the perpendicular magnetic power spectra are found to give a $k^{-5/3}$ behaviour \citep{Telloni}. Despite this limitation, wavelet based analysis of scale-dependent anisotropy can be interesting due to its simplicity and statistical robustness both for the velocity and the magnetic field.\\

Unlike velocity and magnetic field, the anisotropy of electric field fluctuations is not studied extensively. The first (and probably the only till date) study of power and spectral index anisotropy in the electric field power spectra was carried out by projecting the electric field data along and perpendicular to the global mean magnetic field \citep{Mozer}. After following stringent selection criteria, they could find only three suitable intervals from the \textit{THEMIS} and Cluster data. From their analysis, it was found that the parallel power magnitudes were comparable to or greater than the perpendicular powers. Within the frequency range $10^{-2}-1$ Hz, both the spectra had similar shape and inertial range power law slopes of about -5/3. But the results obtained could not be validated over more intervals due to lack of data. In this paper, we investigate the power and spectral index anisotropies of electric field fluctuations with respect to local mean magnetic field. Following the wavelet transform technique as implemented in \citet{Horbury} for magnetic field data, here we decompose a time series of electric field data into wavelet coefficients which are localized in both time and frequency (or wavelet scale). In order to capture the local mean magnetic field, we use Gaussian windows of different standard deviations thereby giving a measure of the corresponding length scale.
\begin{table*}
	\centering
	\caption{Parameter values for the streams analyzed near solar maximum and minimum}
	\label{tab:parameter_table}
	\begin{tabular}{c c c c c c c}
		\hline
		Year & From & To & Solar wind velocity & Proton Density & Proton Parallel Temperature & Proton Perpendicular Temperature \\
		~ & ~ & ~ & $km\ s^{-1}$ & $(n_p)\ cm^{-3}$ & $(T_{\parallel})\ eV$ & $(T_{\perp})\ eV$ \\
		\hline \hline
		2002 & 02 Feb 14:00 & 03 Feb 00:06 & (-393.9, 8.7, -41.7) & 5.7 & 38.9 & 32.2 \\
		2007 & 17 Jan 17:33 & 17 Jan 20:44 & (-623.7, -6.7, -38.8) & 1.04 & 132.8 & 80.6 \\
		\hline
	\end{tabular}
\end{table*}

\section{Data Intervals}
We use the data from Cluster spacecraft (ESA), when it is in the pristine fast solar wind (speed $\sim 550-700 km/s)$ or moderate solar wind (speed $\sim 400-550 km/s)$, outside the earth's magnetic environment. Electric field measurements are made by the electric field and wave (EFW) experiment \citep{Gustafsson} and magnetic field is measured by flux gate magnetometer (FGM) instrument \citep{Balogh} on board the spacecraft. We use the magnetic field and electric field data of cadences 22 and 25 Hz respectively, measured in geocentric solar ecliptic (GSE) coordinate system. Ion moments (velocity, density and temperature) of the solar wind are obtained from cluster ion spectrometry (CIS) experiment \citep{Reme}. All the above data are taken from Cluster 4 (C4) spacecraft. To choose the data intervals for our analysis, first we searched for the streams with longest duration possible (to get the maximum number of data points) with no discontinuities and then checked for the stationarity, which means that the average value of the parameter should change the least during the entire interval. This selection criteria enabled us to choose total 60 intervals for this study with 30 data intervals from the solar activity period (year 2001-2003) and another 30 intervals from the solar calm period (2007) of the solar cycle 23. The selected intervals are from the periods when spacecraft is at geocentric distances of $15 R_E$ to $20 R_E$ with interval lengths lasting from 1 hr to 10 hrs. In addition, during the high solar activity period the data is chosen so as to avoid any transient phenomenon like CMEs (note that the only CME that took place in 2002 was in January). \\

\begin{figure}
	\includegraphics[width=1.0\linewidth]{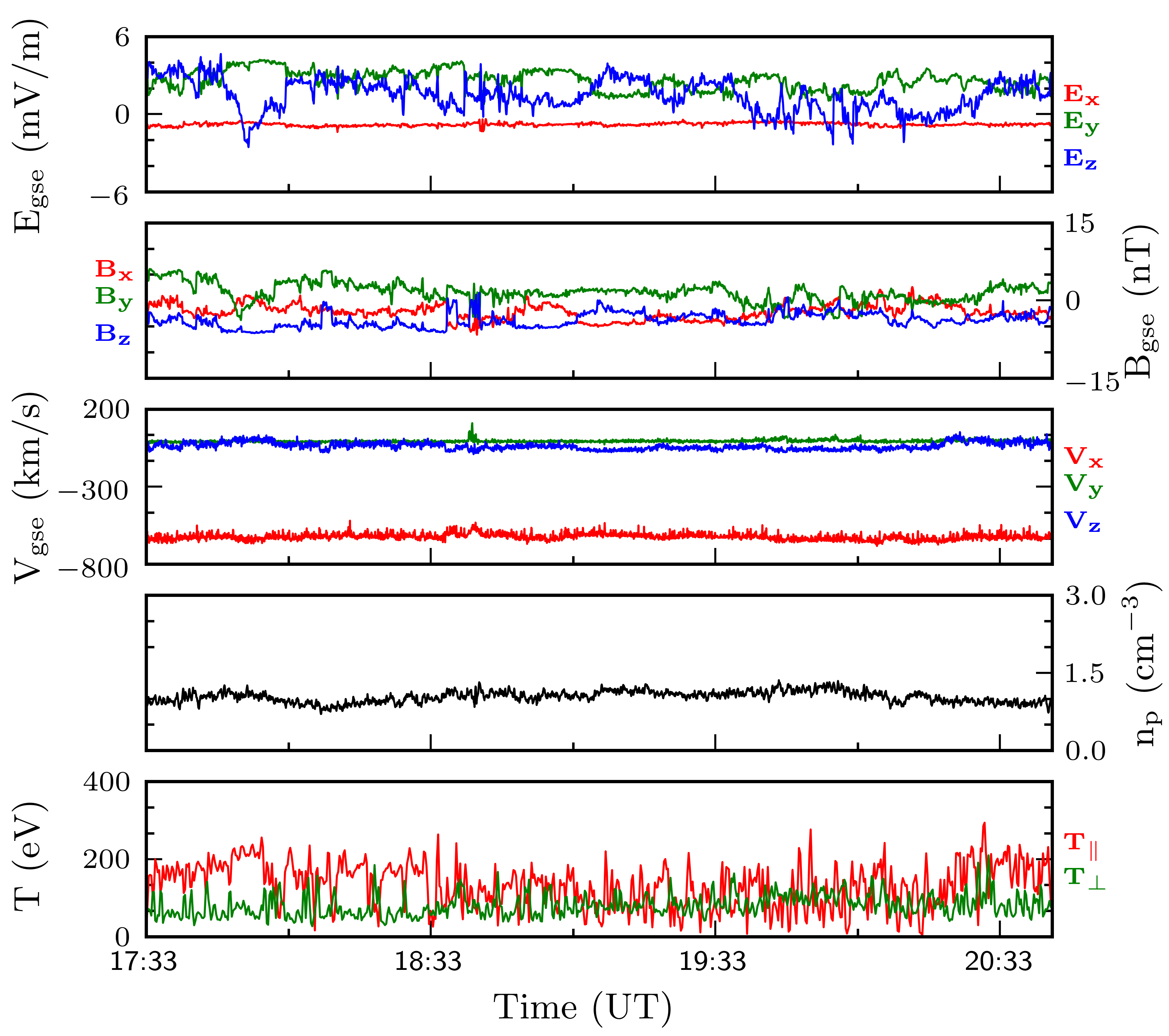}
    \caption{Plasma parameters (proton temperatures, proton density and proton bulk velocity), electric and magnetic field components for the interval of 17 Jan 2007 from 17h33 to 20h44.}
    \label{fig:para}
\end{figure}

In table~\ref{tab:parameter_table}, we mention parameters for two of the longest duration streams that we could find, one each from the period of sun's maximum (year 2002) and minimum activity (year 2007). The spacecraft C4 was at a distance of $18.25 R_e$ and $19 R_e$ from the earth (using ephemeris data), respectively. The proton temperatures $(T_{\parallel})$ and $(T_{\perp})$, proton density $(n_p)$ and proton bulk velocity $(V_x, V_y, V_z)$ values mentioned in the table have been averaged over the respective entire intervals. Fig.~\ref{fig:para} shows the time variation of various plasma properties over the interval from the year 2007. The presented data consists of negligible amount (less than 0.5\%) of data gaps which we have tackled using linear interpolation.
\begin{figure}
\centering
\includegraphics[width=\linewidth]{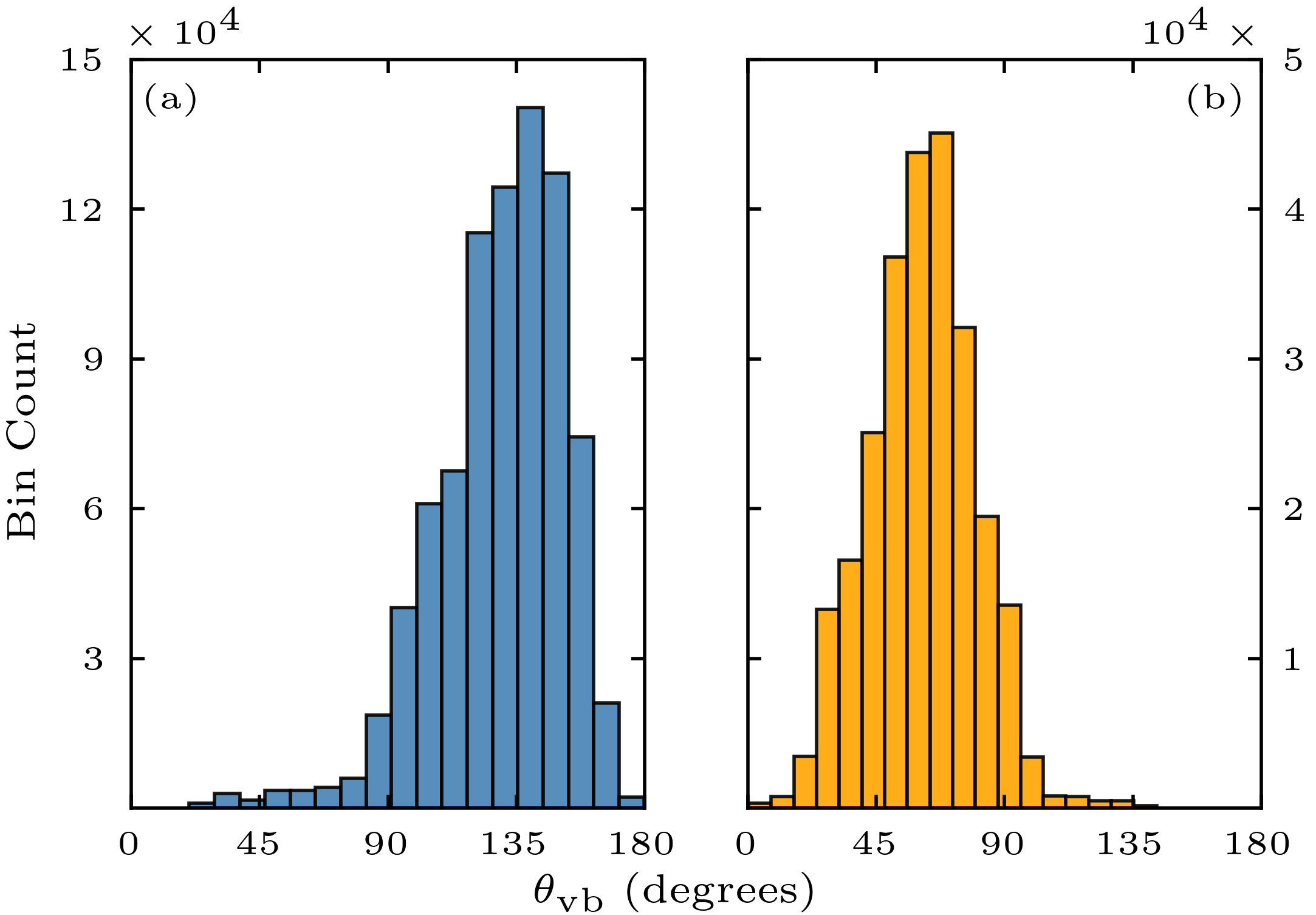}
\caption{Histogram of number of counts for each angle bin computed at the frequency 0.04 Hz (wavelet scale value of 22.59 sec) for $(a)$ the data interval from February 2, 2002 during sun's maximum activity (in blue) and $(b)$ the data interval from January 17, 2007 during solar minimum (in orange).}
\label{fig:hist}
\end{figure}

\section{Analysis Techniques}
In this study, we measure the reduced spectrum which is defined as \citep{Fredricks, Horbury, Papen2015}: 
\begin{equation}
    \mathcal{P}(f) = \displaystyle \int_{}^{} d^3\textbf{k}\,P(\textbf{k})\,\delta(2\pi f - \textbf{k\,.\,V})
\end{equation}

Scale dependent anisotropy of electric field fluctuation spectrum can be measured by studying how its spacecraft-frame power spectrum $\mathcal{P}(f)$ varies with angle between the average flow direction and the LMF. In order to compute the power spectra, 25 Hz electric field data is sub-sampled onto the time tags of 22 Hz magnetic field data by linear interpolation. Wavelet transformation of the electric field data is carried out to decompose the time series into time-frequency space thereby allowing us to calculate the power localised in both time and frequency \citep{Torrence}. For component i of electric field, $E_i(t_k)$, where $t_k = t_0 + k\delta t$, with equal time spacing $\delta t = 1/22 \ s$ and $k=0,1,2...N-1$, the continuous wavelet transform is defined as the convolution of $E_i(t_k)$ with a scaled and translated version of $\psi(\eta)$:
\begin{equation}
    w_i(t_j, f_l) = \displaystyle \sum\limits_{k=0}^{N-1} \ E_i(t_k) \ \psi\left(\frac{t_k - t_j}{s_l}\right)
	\label{eq:wavelet}
\end{equation}
where, $\psi(\eta)$ is called the Morlet wavelet and is given as:
\begin{equation}
  \psi(\eta) = \pi^{-1/4} \ e^{i\omega_0\eta} \ e^{-\eta^{2}/2} \ .
\end{equation}
Here $\eta$ is a non-dimensional time parameter and $\omega_0$ is the non-dimensional frequency taken as 6 for it to construct a nearly orthonormal set of wavelets \citep{Farge}. 
To obtain the set of frequencies for Morlet wavelet, it is convenient to have:
\begin{equation}
    s_m = s_0 \ 2^{m \delta m}, \ m = 0,1,...,M
\end{equation}
where smallest resolvable scale, $s_0 = 2 \ \delta t/(1.03)$. For adequate sampling of scales, we choose $\delta m$ as 0.5 and conduct the analysis for a total of 25 scales. For $\omega_0=6$, the wavelet scale $s_l$ is related to the frequency $f_l$ as $f_l=0.97/s_l$ from which we get a set of 25 frequencies. Instead of computing the wavelet coefficients from equation~(\ref{eq:wavelet}), we calculate them in Fourier space using fast Fourier transform (FFT) and convolution theorem, which is considered to be faster. We calculate the wavelet coefficients at 25 frequencies ranging from 2.7 mHz to 11 Hz (Nyquist frequency for the data used). Power in component i at time $t_j$ and frequency $f_l$ is defined as: 
\begin{equation}
    \mathcal{P}_{ii}(t_j, f_l) \displaystyle \propto  \ \frac{2\delta t}{N} \ \mid w_i(t_j, f_l) \mid^2 \ .
\end{equation}
For the vector field \textbf{E}(t), the trace power is the sum of the power of the three orthogonal components, such that:\\ 
Trace power, $\mathcal{P} = \mathcal{P}_{xx} + \mathcal{P}_{yy} + \mathcal{P}_{zz}$ . We now calculate the scale dependent mean magnetic field or the LMF at time $t_j$ and wavelet scale $s_l$. This is obtained by modulating $B_i$, the ith component of the magnetic field time series, with a Gaussian curve centered at time $t_j$ and scaled with $s_l$, and is given by:
\begin{equation}
    \bar{b}_i(t_j, s_l) = \displaystyle \sum\limits_{k=0}^{N-1} \ B_i(t_k) \ \exp\left[-\frac{(t_k-t_j)^2}{2s_l^2}\right] \ .
\end{equation}
This results in a time series of vectors $\bar{\textbf{b}}(t_j, s_l)$ for each scale $s_l$ pointing in the direction of LMF.  The inclination ($\theta_{vb}$, again a scale-dependent quantity) of this mean field with respect to the average flow direction $\textbf{V}_{sw}$ of solar wind (sw) can be given by:
\begin{equation}
     \theta_{vb} (t_j, s_l) = \displaystyle \cos^{-1} \left(\frac{\textbf{V}_{sw}\cdot \ \bar{\textbf{b}}}{\mid\textbf{V}_{sw}\mid \ \mid\bar{\textbf{b}}\mid} \right).
	\label{eq:angle}
\end{equation}
\color{black}
The inertial range turbulence is found to be axisymmetric about the magnetic field direction \textit{i.e.} the power spectra are independent of the azimuthal angle $\phi$ and hence we consider the power values averaged over all $\phi$ \citep{Horbury}. Upon obtaining the directions of the local mean magnetic field, we now bin the powers of each scale as a function of $\theta_{vb}$. We construct bins of equal spacing of $10^\circ$. As it is evident from the Fig.~\ref{fig:hist}, for the 2002 interval, we get 16 bins ($20^\circ$ to $180^\circ$) and for that of 2007, we have only 14 bins ($0^\circ$ to $140^\circ$). The distribution of bin counts are found to be almost similar at all frequencies from 2.7 mHz to 11 Hz. For each bin and for a given frequency $f_l$, we average the trace power values corresponding to all the angles that lie in the bin. By doing this, we obtain the average trace power value $\mathcal{P}(f_l,\,\theta _{vb})$ corresponding to the frequency $f_l$ for that particular angle bin. For the sake of reasonable statistical sampling, we reject those angle bins which have less than 100 contributing power levels.
\begin{figure*}
	\includegraphics[width=\linewidth]{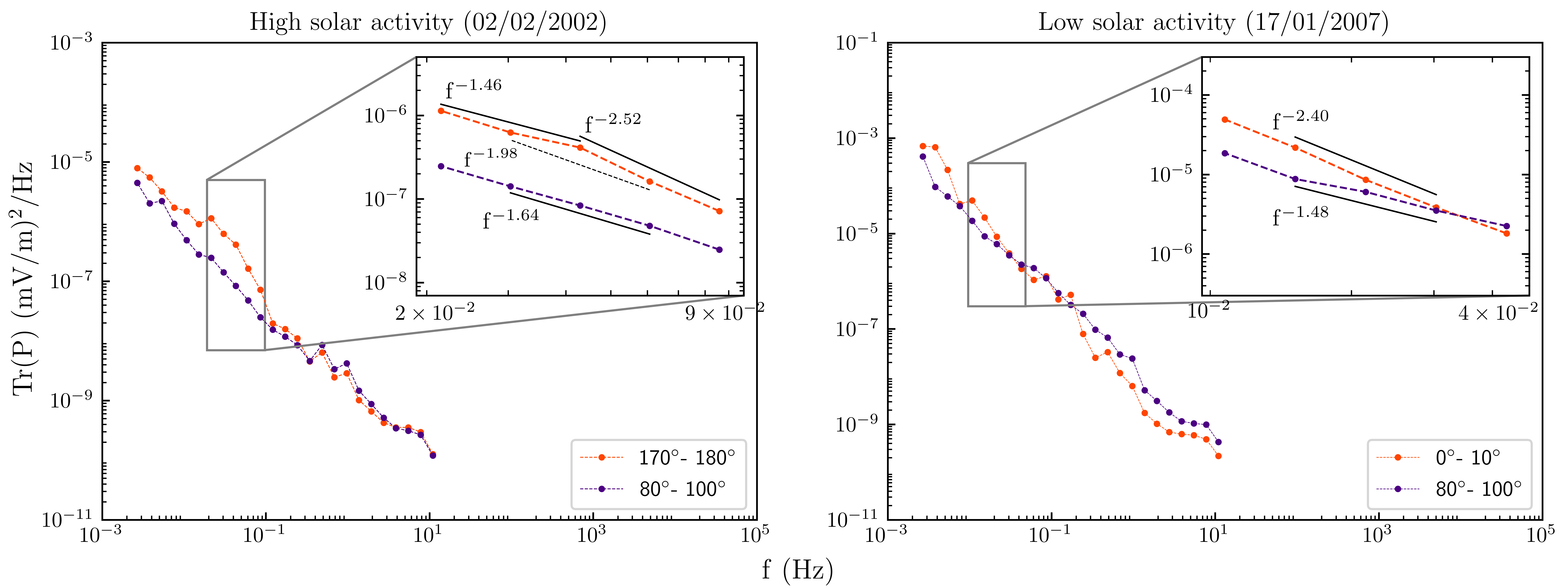}
    \caption{Electric power spectra at two different angle ranges of the local mean magnetic field to the flow direction: $80^\circ$-$100^\circ$ (blue) and $170^\circ$-$180^\circ$ (orange) for the data interval 02 Feb, 2002 14:00 to 03 Feb, 2002 00:06 (solar activity) whereas $80^\circ$-$100^\circ$ (blue) and $0^\circ$-$10^\circ$ (orange) for the data interval 17 Jan, 2007 17:33 to 20:44 (solar calm) }
    \label{fig:ladder_plot}
\end{figure*}
Of the 60 intervals used in this study, only for 4 intervals, we found contribution from both the parallel (or antiparallel) and perpendicular direction with respect to the LMF. Finally, in this paper, we have chosen to present the study of two intervals, one from each of the years 2002 and 2007.

\section{Results}
Fig.~\ref{fig:ladder_plot} shows, for two aforementioned intervals, the turbulent power spectra of electric field fluctuations along and perpendicular to the LMF. From the figures it is evident that unlike the previous studies \citep{Mozer}, here we obtain anisotropies in both power levels and spectral indices. In particular, the nature of anisotropy is found to depend strongly on the frequency. In both intervals, clear discrepancy between the parallel (or the antiparallel) and the perpendicular spectra is observed in the upper MHD frequency range \textit{i.e.} from $10^{-2}-10^{-1}$ Hz. Most of the time, it is found that the parallel power is larger and steeper than the perpendicular power. For 2002, where the wind is moderately slow, the antiparallel electric power is found to be roughly 4 times the perpendicular power. The antiparallel and the perpendicular spectral slopes are obtained as $-1.98$ and $-1.64$, which looks very similar to that of magnetic power anisotropy which is consistent with critical balance theory. However here we have further subdivided the parallel spectra into a very flat ($-1.46$) and a very steep ($-2.52$) part which give together a slope of nearly $-1.98$. During the interval of 2007, the MHD range anisotropy also represents an initial dominance of parallel electric power over the perpendicular one (by a factor of around 2.5). The parallel and perpendicular spectral indices are $-2.4$ and $-1.48$ respectively. Note that, the electric field amplitudes were much stronger in 2007 and hence a common factor of around 10 is observed between the power amplitudes of 2002 and 2007. After this range of prominent anisotropy, the parallel and perpendicular spectra behave randomly for different intervals of our study, therefore making it difficult to interpret. However, beyond 1 Hz, both the spectra get flat possibly due to the instrumental noise floor (which lies around $\sim 10^{-10} (mV/m)^2/Hz$) \citep{Salem}.
\begin{figure}
	\centering
	\includegraphics[width=\linewidth]{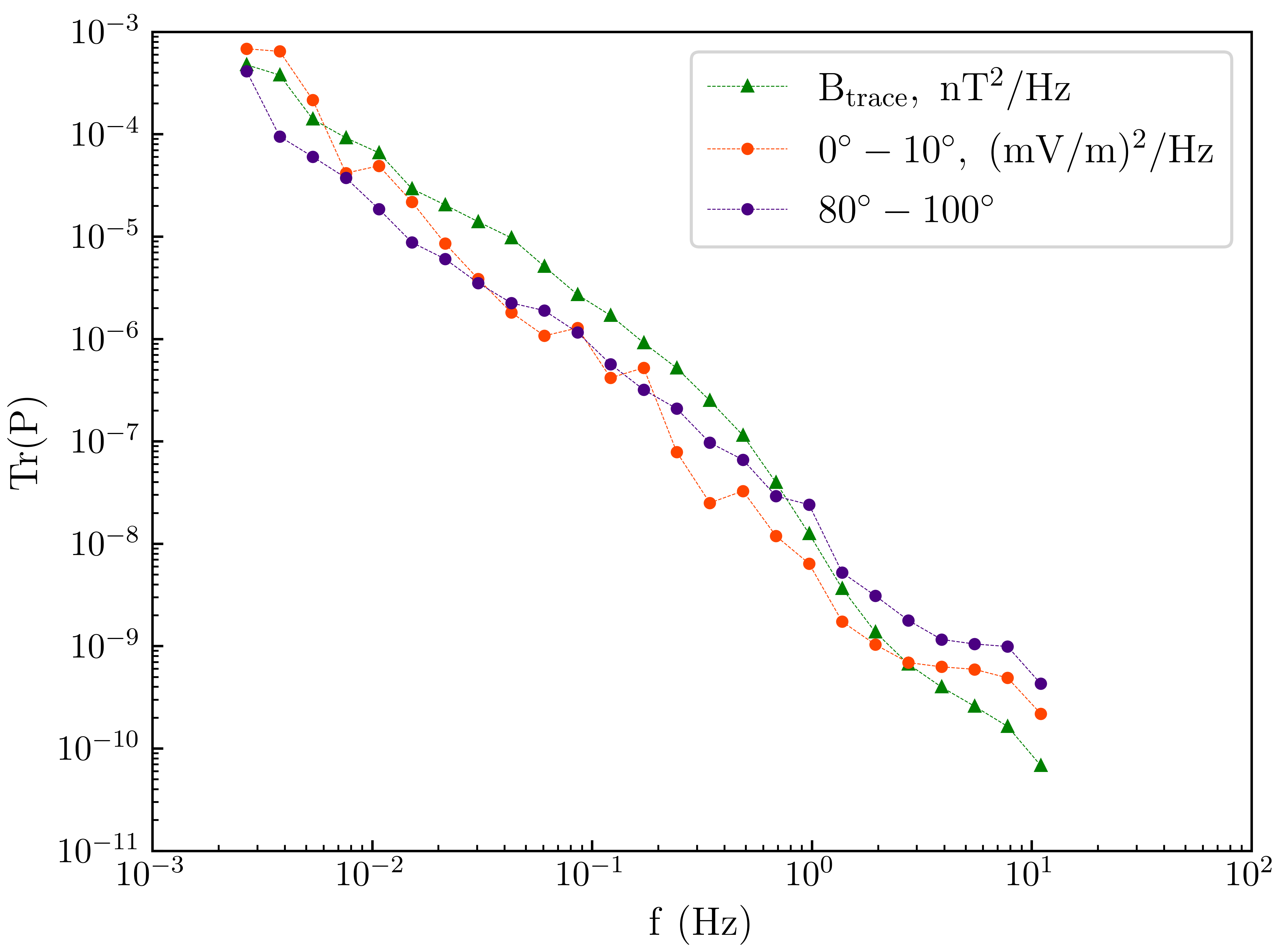}
    \caption{Perpendicular trace electric power spectrum (flow direction to local magnetic field angle ranging from $80^\circ$-\ $100^\circ$) in blue, Parallel electric fluctuations spectrum (angle ranging from $0^\circ$-\ $10^\circ$) in orange and trace magnetic field spectrum (in green) plotted with respect to frequency for the data interval 17 Jan, 2007.}
    \label{fig:comparison_plot}
\end{figure}
\\

To observe how the magnetic power spectrum scales with respect to that of parallel and perpendicular electric field fluctuations, we plot the three spectra together for the 2007 interval, as shown in Fig.~\ref{fig:comparison_plot}. It can be seen that the trace power spectrum of magnetic field fluctuations follows the perpendicular electric power spectrum upto the frequency of approximately $0.3$ Hz. Quantitatively, the spectral indices are measured to be $-1.54$ for the former and $-1.43$ for the latter in the range of frequencies $0.01 - 0.3$ Hz. After around $\sim 0.3$ Hz, magnetic field spectrum becomes steeper via smooth transition unlike both the electric field spectra. Furthermore, it can be seen that both the magnetic and perpendicular electric fluctuations spectrum have a breakpoint near $\sim 0.6$ Hz after which the magnetic spectrum becomes more steeper than the electric one. Due to the unavailability of high cadence data, it is difficult to conclude reasonably for the frequency range above 1 Hz \citep{Mozer}. In the discussion section of \citep{Mozer}, they state that the parallel electric field spectrum coincides with trace magnetic field spectrum which is not expected for Alfv\'enic turbulence in the inertial range. They observe identical slopes for both the magnetic and parallel electric field spectra possibly because of using the global mean magnetic field and not the LMF. Interestingly, with the use of local mean magnetic field in our study, we find that parallel electric spectrum does not coincide with that of magnetic field but the perpendicular spectrum, which is normally expected for Alfv\'enic turbulence in the solar wind \citep{Bianetal}.
\begin{figure*}
	\includegraphics[width=\linewidth]{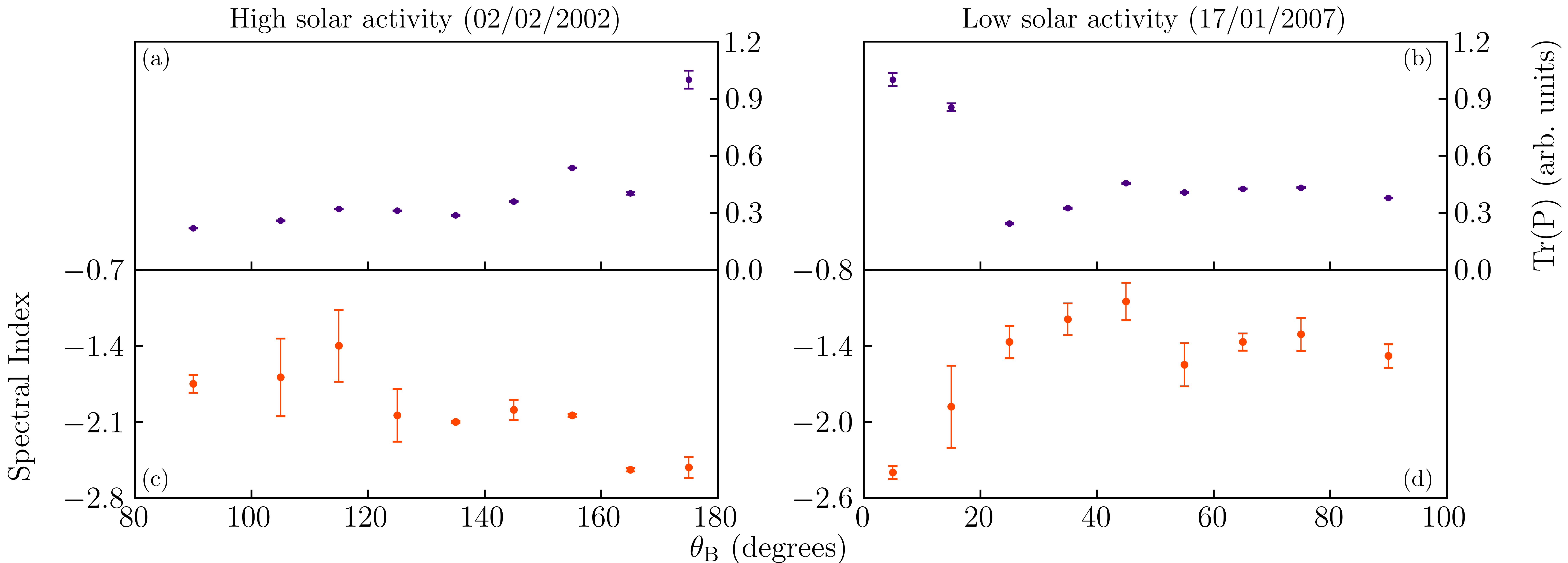}
    \caption{(a), (b): Variation of average trace powers of electric field (in blue) with the angle between average flow and local mean magnetic field for both the data intervals from 2002 (at frequency 21 mHz) and 2007 (at frequency 11 mHz) respectively. (c), (d): Power law indices (in orange) plotted against the flow-to-field angles, fitted for frequency range $0.043 - 0.086$ Hz for 2002 data and frequency range $0.011 - 0.043$ Hz for 2007 data interval.}
    \label{fig:si_tr}
\end{figure*}

Next, we present the average electric powers (normalized by parallel power) and spectral indices as a function of angles between the average flow direction and the local mean magnetic field. Following \citet{Wicks11}, we calculate the errors in average powers from the respective standard deviation and those in spectral indices from the errors introduced by least square fitting. For the interval from 02 February 2002 (Fig.~\ref{fig:si_tr} (a)), it is clearly evident that the power in antiparallel electric fluctuations is approximately 4 times more than that in the perpendicular direction. Similarly, for the data interval from 17 January, 2007 (Fig.~\ref{fig:si_tr} (b)) the parallel power is found to be greater than the perpendicular one. Interestingly, we can see nearly similar behaviour for the angle variation of trace powers during the high solar activity and low solar activity but the shape of these variations is different from that obtained for magnetic field fluctuations as reported in \citet{Horbury}.However the variation of spectral indices with angles is found to follow roughly the same nature as reported by \citet{Horbury}. \citet{Mozer} reported in their paper that the parallel electric powers can be greater than or comparable to the perpendicular powers. But from our scale dependent analysis of both the intervals, we can firmly say that the parallel electric powers are greater than the perpendicular powers. As for the spectral index anisotropy, shown in Fig.~\ref{fig:si_tr} (c), (d) the parallel (or antiparallel) electric field spectrum is found to be steeper than the perpendicular spectrum for both the intervals from two different periods of solar cycle. Qualitatively, this feature of power laws of electric fluctuations spectrum in two different angle ranges is similar to that found for magnetic fluctuations spectrum. Out of two types of anisotropies studied in this paper for electric field fluctuations, it is interesting to note that the power anisotropy is opposite and spectral index anisotropy is similar to what has been found for magnetic field fluctuations. 

To conclude, our study clearly shows that (i) electric fluctuations spectrum undergoes prominent power and spectral index anisotropies when the analysis is done with respect to the local mean field, (ii) unlike \citet{Mozer}, the parallel electric power spectrum does not share the same slope as that of the magnetic power spectrum thereby indicating the clear signature of Alfv\'enic turbulence \citep{Bianetal}. \citet{Bianetal} (and with in references) also mention that the presence of the parallel electric field produced by Alfv\'enic turbulence and its further study through theory and observations might be helpful for understanding the turbulent acceleration of particles through wave-particle interactions. \citet{Kellogg} mentioned that diffusion due to the electric fields is the dominant process resulting in the isotropy of the proton distributions. Hence, the study of spectral structure of electric field in direction parallel and perpendicular to mean magnetic field could also shed some light on the particle distribution in space. With the launch of future missions and the availability of higher resolution data, it will be possible to probe into the anisotropy in the smaller scales. Another possible future work can consist of the use and analysis of the multi-spacecraft data of Cluster to study the anisotropy of electric power using the structure functions method following \citet{Chen2011}.

\section*{Acknowledgements}
This research has been supported by the DST INSPIRE research grant (DST/PHY/2017514).  

\section*{Data Availability}
The data used for this study is available at the Cluster Science Archive (\url{https://csa.esac.esa.int/csa/aio/html/home_main.shtml}). 

\bibliographystyle{mnras}
\bibliography{MNRAS} 

\providecommand{\noopsort}[1]{}\providecommand{\singleletter}[1]{#1}%
\begin{thebibliography}{}
\makeatletter
\relax
\def\mn@urlcharsother{\let\do\@makeother \do\$\do\&\do\#\do\^\do\_\do\%\do\~}
\def\mn@doi{\begingroup\mn@urlcharsother \@ifnextchar [ {\mn@doi@}
  {\mn@doi@[]}}
\def\mn@doi@[#1]#2{\def\@tempa{#1}\ifx\@tempa\@empty \href
  {http://dx.doi.org/#2} {doi:#2}\else \href {http://dx.doi.org/#2} {#1}\fi
  \endgroup}
\def\mn@eprint#1#2{\mn@eprint@#1:#2::\@nil}
\def\mn@eprint@arXiv#1{\href {http://arxiv.org/abs/#1} {{\tt arXiv:#1}}}
\def\mn@eprint@dblp#1{\href {http://dblp.uni-trier.de/rec/bibtex/#1.xml}
  {dblp:#1}}
\def\mn@eprint@#1:#2:#3:#4\@nil{\def\@tempa {#1}\def\@tempb {#2}\def\@tempc
  {#3}\ifx \@tempc \@empty \let \@tempc \@tempb \let \@tempb \@tempa \fi \ifx
  \@tempb \@empty \def\@tempb {arXiv}\fi \@ifundefined
  {mn@eprint@\@tempb}{\@tempb:\@tempc}{\expandafter \expandafter \csname
  mn@eprint@\@tempb\endcsname \expandafter{\@tempc}}}

\bibitem[\protect\citeauthoryear{Alexandrova, Saur, Lacombe, Mangeney,
  Mitchell, Schwartz  \& Robert}{Alexandrova et~al.}{2009}]{Alexandrova2009}
Alexandrova O.,  Saur J.,  Lacombe C.,  Mangeney A.,  Mitchell J.,  Schwartz
  S.~J.,   Robert P.,  2009, \mn@doi [Phys. Rev. Lett.]
  {10.1103/PhysRevLett.103.165003}, 103, 165003

\bibitem[\protect\citeauthoryear{{Bale}, {Kellogg}, {Mozer}, {Horbury}  \&
  {Reme}}{{Bale} et~al.}{2005}]{Bale}
{Bale} S.~D.,  {Kellogg} P.~J.,  {Mozer} F.~S.,  {Horbury} T.~S.,   {Reme} H.,
  2005, \mn@doi [\prl] {10.1103/PhysRevLett.94.215002}, \href
  {https://ui.adsabs.harvard.edu/abs/2005PhRvL..94u5002B} {94, 215002}

\bibitem[\protect\citeauthoryear{{Balogh} et~al.,}{{Balogh}
  et~al.}{1997}]{Balogh}
{Balogh} A.,  et~al., 1997, \mn@doi [\ssr] {10.1023/A:1004970907748}, \href
  {https://ui.adsabs.harvard.edu/abs/1997SSRv...79...65B} {79, 65}

\bibitem[\protect\citeauthoryear{Banerjee}{Banerjee}{2014}]{banerjeethese}
Banerjee S.,  2014, Theses, {Universit{\'e} Paris Sud - Paris XI}, \url
  {https://tel.archives-ouvertes.fr/tel-01087024}

\bibitem[\protect\citeauthoryear{{Bian}, {Kontar}  \& {Brown}}{{Bian}
  et~al.}{2010}]{Bianetal}
{Bian} N.~H.,  {Kontar} E.~P.,   {Brown} J.~C.,  2010, \mn@doi [\aap]
  {10.1051/0004-6361/201014048}, \href
  {https://ui.adsabs.harvard.edu/abs/2010A&A...519A.114B} {519, A114}

\bibitem[\protect\citeauthoryear{Chen, Horbury, Schekochihin, Wicks,
  Alexandrova  \& Mitchell}{Chen et~al.}{2010}]{Chen2010}
Chen C. H.~K.,  Horbury T.~S.,  Schekochihin A.~A.,  Wicks R.~T.,  Alexandrova
  O.,   Mitchell J.,  2010, \mn@doi [Phys. Rev. Lett.]
  {10.1103/PhysRevLett.104.255002}, 104, 255002

\bibitem[\protect\citeauthoryear{Chen, Mallet, Yousef, Schekochihin  \&
  Horbury}{Chen et~al.}{2011}]{Chen2011}
Chen C. H.~K.,  Mallet A.,  Yousef T.~A.,  Schekochihin A.~A.,   Horbury T.~S.,
   2011, \mn@doi [Monthly Notices of the Royal Astronomical Society]
  {10.1111/j.1365-2966.2011.18933.x}, 415, 3219

\bibitem[\protect\citeauthoryear{Cho \& Vishniac}{Cho \&
  Vishniac}{2000}]{Cho2000}
Cho J.,  Vishniac E.~T.,  2000, \mn@doi [The Astrophysical Journal]
  {10.1086/309213}, 539, 273

\bibitem[\protect\citeauthoryear{{Farge}}{{Farge}}{1992}]{Farge}
{Farge} M.,  1992, \mn@doi [Annual Review of Fluid Mechanics]
  {10.1146/annurev.fl.24.010192.002143}, \href
  {https://ui.adsabs.harvard.edu/abs/1992AnRFM..24..395F} {24, 395}

\bibitem[\protect\citeauthoryear{{Fredricks} \& {Coroniti}}{{Fredricks} \&
  {Coroniti}}{1976}]{Fredricks}
{Fredricks} R.~W.,  {Coroniti} F.~V.,  1976, \mn@doi [\jgr]
  {10.1029/JA081i031p05591}, \href
  {https://ui.adsabs.harvard.edu/abs/1976JGR....81.5591F} {81, 5591}

\bibitem[\protect\citeauthoryear{{Goldreich} \& {Sridhar}}{{Goldreich} \&
  {Sridhar}}{1995}]{Goldreich}
{Goldreich} P.,  {Sridhar} S.,  1995, \mn@doi [\apj] {10.1086/175121}, \href
  {https://ui.adsabs.harvard.edu/abs/1995ApJ...438..763G} {438, 763}

\bibitem[\protect\citeauthoryear{{Gustafsson} et~al.,}{{Gustafsson}
  et~al.}{1997}]{Gustafsson}
{Gustafsson} G.,  et~al., 1997, \mn@doi [\ssr] {10.1023/A:1004975108657}, \href
  {https://ui.adsabs.harvard.edu/abs/1997SSRv...79..137G} {79, 137}

\bibitem[\protect\citeauthoryear{Horbury, Balogh, Forsyth  \& Smith}{Horbury
  et~al.}{1995}]{Horbury95}
Horbury T.~S.,  Balogh A.,  Forsyth R.~J.,   Smith E.~J.,  1995, \mn@doi
  [Geophysical Research Letters] {10.1029/95GL03012}, 22, 3405

\bibitem[\protect\citeauthoryear{{Horbury}, {Forman}  \& {Oughton}}{{Horbury}
  et~al.}{2008}]{Horbury}
{Horbury} T.~S.,  {Forman} M.,   {Oughton} S.,  2008, \mn@doi [\prl]
  {10.1103/PhysRevLett.101.175005}, \href
  {https://ui.adsabs.harvard.edu/abs/2008PhRvL.101q5005H} {101, 175005}

\bibitem[\protect\citeauthoryear{{Kellogg}, {Bale}, {Mozer}, {Horbury}  \&
  {Reme}}{{Kellogg} et~al.}{2006}]{Kellogg}
{Kellogg} P.~J.,  {Bale} S.~D.,  {Mozer} F.~S.,  {Horbury} T.~S.,   {Reme} H.,
  2006, \mn@doi [\apj] {10.1086/499265}, \href
  {https://ui.adsabs.harvard.edu/abs/2006ApJ...645..704K} {645, 704}

\bibitem[\protect\citeauthoryear{Milano, Matthaeus, Dmitruk  \&
  Montgomery}{Milano et~al.}{2001}]{Milano2001}
Milano L.~J.,  Matthaeus W.~H.,  Dmitruk P.,   Montgomery D.~C.,  2001, \mn@doi
  [Physics of Plasmas] {10.1063/1.1369658}, 8, 2673

\bibitem[\protect\citeauthoryear{{Mozer} \& {Chen}}{{Mozer} \&
  {Chen}}{2013}]{Mozer}
{Mozer} F.~S.,  {Chen} C.~H.~K.,  2013, \mn@doi [\apjl]
  {10.1088/2041-8205/768/1/L10}, \href
  {https://ui.adsabs.harvard.edu/abs/2013ApJ...768L..10M} {768, L10}

\bibitem[\protect\citeauthoryear{Oughton \& Matthaeus}{Oughton \&
  Matthaeus}{2020}]{Oughton}
Oughton S.,  Matthaeus W.~H.,  2020, \mn@doi [The Astrophysical Journal]
  {10.3847/1538-4357/ab8f2a}, 897, 37

\bibitem[\protect\citeauthoryear{Podesta}{Podesta}{2009}]{Podesta2009}
Podesta J.~J.,  2009, \mn@doi [The Astrophysical Journal]
  {10.1088/0004-637x/698/2/986}, 698, 986

\bibitem[\protect\citeauthoryear{Podesta, Roberts  \& Goldstein}{Podesta
  et~al.}{2006}]{Podesta2006}
Podesta J.~J.,  Roberts D.~A.,   Goldstein M.~L.,  2006, \mn@doi [Journal of
  Geophysical Research: Space Physics] {10.1029/2006JA011834}, 111

\bibitem[\protect\citeauthoryear{{Reme} et~al.,}{{Reme} et~al.}{1997}]{Reme}
{Reme} H.,  et~al., 1997, \mn@doi [\ssr] {10.1023/A:1004929816409}, \href
  {https://ui.adsabs.harvard.edu/abs/1997SSRv...79..303R} {79, 303}

\bibitem[\protect\citeauthoryear{Sahraoui, Goldstein, Robert  \&
  Khotyaintsev}{Sahraoui et~al.}{2009}]{Sahraoui2009}
Sahraoui F.,  Goldstein M.~L.,  Robert P.,   Khotyaintsev Y.~V.,  2009, \mn@doi
  [Phys. Rev. Lett.] {10.1103/PhysRevLett.102.231102}, 102, 231102

\bibitem[\protect\citeauthoryear{{Salem}, {Howes}, {Sundkvist}, {Bale},
  {Chaston}, {Chen}  \& {Mozer}}{{Salem} et~al.}{2012}]{Salem}
{Salem} C.~S.,  {Howes} G.~G.,  {Sundkvist} D.,  {Bale} S.~D.,  {Chaston}
  C.~C.,  {Chen} C.~H.~K.,   {Mozer} F.~S.,  2012, \mn@doi [\apjl]
  {10.1088/2041-8205/745/1/L9}, \href
  {https://ui.adsabs.harvard.edu/abs/2012ApJ...745L...9S} {745, L9}

\bibitem[\protect\citeauthoryear{{Sari} \& {Valley}}{{Sari} \&
  {Valley}}{1976}]{Sari}
{Sari} J.~W.,  {Valley} G.~C.,  1976, \mn@doi [\jgr] {10.1029/JA081i031p05489},
  \href {https://ui.adsabs.harvard.edu/abs/1976JGR....81.5489S} {81, 5489}

\bibitem[\protect\citeauthoryear{{Taylor}}{{Taylor}}{1938}]{Taylor}
{Taylor} G.~I.,  1938, \mn@doi [Proceedings of the Royal Society of London
  Series A] {10.1098/rspa.1938.0032}, \href
  {https://ui.adsabs.harvard.edu/abs/1938RSPSA.164..476T} {164, 476}

\bibitem[\protect\citeauthoryear{Telloni, Carbone, Bruno, Sorriso-Valvo, Zank,
  Adhikari  \& Hunana}{Telloni et~al.}{2019}]{Telloni}
Telloni D.,  Carbone F.,  Bruno R.,  Sorriso-Valvo L.,  Zank G.~P.,  Adhikari
  L.,   Hunana P.,  2019, \mn@doi [The Astrophysical Journal]
  {10.3847/1538-4357/ab517b}, 887, 160

\bibitem[\protect\citeauthoryear{{Tessein}, {Smith}, {MacBride}, {Matthaeus},
  {Forman}  \& {Borovsky}}{{Tessein} et~al.}{2009}]{Tessein}
{Tessein} J.~A.,  {Smith} C.~W.,  {MacBride} B.~T.,  {Matthaeus} W.~H.,
  {Forman} M.~A.,   {Borovsky} J.~E.,  2009, \mn@doi [\apj]
  {10.1088/0004-637X/692/1/684}, \href
  {https://ui.adsabs.harvard.edu/abs/2009ApJ...692..684T} {692, 684}

\bibitem[\protect\citeauthoryear{{Torrence} \& {Compo}}{{Torrence} \&
  {Compo}}{1998}]{Torrence}
{Torrence} C.,  {Compo} G.~P.,  1998, \mn@doi [Bulletin of the American
  Meteorological Society] {10.1175/1520-0477(1998)079<0061:APGTWA>2.0.CO;2},
  \href {https://ui.adsabs.harvard.edu/abs/1998BAMS...79...61T} {79, 61}

\bibitem[\protect\citeauthoryear{Wicks, Horbury, Chen  \& Schekochihin}{Wicks
  et~al.}{2011}]{Wicks11}
Wicks R.~T.,  Horbury T.~S.,  Chen C. H.~K.,   Schekochihin A.~A.,  2011,
  \mn@doi [Phys. Rev. Lett.] {10.1103/PhysRevLett.106.045001}, 106, 045001

\bibitem[\protect\citeauthoryear{von Papen \& Saur}{von Papen \&
  Saur}{2015}]{Papen2015}
von Papen M.,  Saur J.,  2015, \mn@doi [The Astrophysical Journal]
  {10.1088/0004-637x/806/1/116}, 806, 116

\makeatother
\end{thebibliography}

\bsp	
\label{lastpage}
\end{document}